\documentclass[useAMS,usegraphicx]{mn2e}
\usepackage{rotate}
\usepackage{times}
\newif\ifAMStwofonts
\AMStwofontstrue
\def\gs{\mathrel{\hbox{\rlap{\hbox{\lower4pt\hbox{$\sim$}}}\hbox{$>$}}}}
\def\ls{\mathrel{\hbox{\rlap{\hbox{\lower4pt\hbox{$\sim$}}}\hbox{$<$}}}}

\def\Msun{M$_{\odot}$}

\def\einstein{{\it Einstein}}
\def\exosat{{\it EXOSAT}}
\def\chandra{{\it Chandra}}

\def\hst{{\it HST}}
\def\fuse{{\it FUSE}}
\def\rosat{{\it ROSAT}}

\def\heao{{\it HEAO-1} A2}
\def\asca{{\it ASCA}}

\def\xte{{\it RXTE}}
\def\xmm{{\it XMM-Newton}}
\def\ginga{{\it Ginga}}

\def\et{{et al.\ }}
\def\ark{{Ark~120}}
\def\3c{{3C~273}}

\def\arcs{{\hbox{$^{\prime\prime}$}}}
\def\deg{^{\circ}}
\def\A{{\rm\thinspace \AA}}

\def\keV{{\rm\thinspace keV}}

\def\Msun{\hbox{$\rm\thinspace M_{\odot}$}}

\def\rg{\hbox{$r_{\rm g}$}}
\def\rin{\hbox{$r_{\rm in}$}}
\def\rout{\hbox{$r_{\rm out}$}}

\title[An \xmm\ observation of \ark]
      {An \xmm\ observation of \ark: the X-ray spectrum of a `bare'
      Seyfert 1 nucleus}

\author[Vaughan \et]
       {S. Vaughan,$^{1}$
        A. C. Fabian,$^{2}$
        D. R. Ballantyne,$^{3}$
        A. De Rosa,$^{4}$ 
        L. Piro$^{4}$ and
        G. Matt$^{5}$\\
$^{1}$X-Ray and Observational Astronomy Group, Department of Physics
       and Astronomy, University of Leicester, Leicester, LE1 7RH\\
$^{2}$Institute of Astronomy, Madingley Road, Cambridge, CB3 0HA\\
$^{3}$Canadian Institute for Theoretical Astrophysics, McLennan Labs, 60 St. George Street, Toronto, Ontario M5S 3H8, Canada\\
$^{4}$Istituto di Astrofisica Spaziale, C.N.R.,Via Fosso del Cavaliere, Roma, Italy\\
$^{5}$Dipartimento di Fisica, Universit\`{a} degli Studi ``Roma Tre,'' Via della Vasca Navale 84, 00146 Roma, Italy
}

\date{Accepted: 25/2/2004; submitted: 6/2/2004; in original form: 24/12/2003}
\pagerange{\pageref{firstpage}--\pageref{lastpage}}

\pubyear{2004}
\begin{document}
\maketitle
\label{firstpage}

\begin{abstract}
We report on a long ($100$~ks) \xmm\ observation of the bright Seyfert
1 galaxy Arakelian 120.  The source previously showed no signs of intrinsic
reddening in its infrared--ultraviolet continuum and previous
observations had shown no evidence for ionized absorption in either the
ultraviolet or X-ray bands.  The new \xmm\ RGS data place tight limits on
the presence of an ionized X-ray absorber and confirm that the X-ray
spectrum of \ark\ is essentially unmodified by intervening matter.
Thus \ark\ can be considered a `bare' Seyfert 1 nucleus.
This observation therefore offers a  clean view of the X-ray spectrum of a
`normal' Seyfert galaxy free from absorption effects. 
The spectrum shows a 
Doppler broadened iron emission  line ($FWHM\sim 3\times 10^4$~km
s$^{-1}$) and a smooth, continuous soft excess which appears to peak at
an energy $\approx 0.5$~keV.  This adds weight to the claim that
genuine soft excesses  (i.e. those due to a real steepening of the
underlying continuum below $\sim 2$~keV) are ubiquitous
in Seyfert 1 spectra.  However, the detailed shape of the
excess could not be reproduced by any of the simple models tested
(power-laws, blackbodies, Comptonised blackbodies, accretion disc
reflection).  This observation therefore demonstrates both the need to
understand the soft excess (as a significant contributor to the
luminosity of most Seyfert 1s) and the inability of the existing,
simple models to explain it.
\end{abstract}

\begin{keywords}
galaxies: active -- 
galaxies: nuclei --
galaxies: Seyfert -- 
X-rays: galaxies 
\end{keywords}

\section{Introduction}
\label{sect:intro}

Arakelian 120 (aka Mrk 1095) is a luminous Seyfert 1 galaxy at a
redshift $z=0.0323$. It was the subject of an early attempt at
reverberation mapping, and was important in demonstrating the compact
size of the optical broad line region (Peterson \et 1985; Peterson \&
Gaskell 1991).  More recent optical monitoring campaigns have yielded
an estimate of the mass of the central black hole of $\sim 2 \times
10^{8}$~\Msun (Wandel, Peterson \& Malkan 1999).  The bolometric
luminosity for the nucleus is $L_{\rm bol} \gs 
10^{45}$ erg s$^{-1}$ (Edelson \& Malkan 1996 estimated the total
$0.1-100~\mu{\rm m}$ luminosity to be $\approx 8\times 10^{44}$  erg
s$^{-1}$).  This would suggest it is radiating at $L/L_{\rm Edd} \gs
0.05$, where $L_{\rm Edd}$ is the Eddington luminosity for a  $2
\times10^{8}$~\Msun\ black hole. The nucleus is radio-quiet  but shows
a slight extension in its radio image (Condon \et 1998; Ho 2002).
Ward \et (1987) used broad-band photometry to identify \ark\ as a
`bare' Seyfert nucleus, i.e. one free from significant reddening or
contamination from the host galaxy.  The host is a low-inclination
spiral galaxy (Hubble type S0/a, inclination $i \approx 26\deg$;
Nordgren \et 1995) .

\ark\ has been observed with most of the major X-ray observatories.
An \exosat\ observation showed \ark\ to have a steep soft X-ray
spectrum (Turner \& Pounds 1989), as did a subsequent \rosat\
observation (Brandt \et 1993). Furthermore, these X-ray observations
showed no indication of any `warm absorption' features -- i.e.
discrete absorption features often found in the soft X-ray band and
caused by absorption in photoionized gas along the line-of-sight to
the nucleus. Warm absorption systems are common in Seyfert 1s
(Reynolds 1997; Crenshaw, Kraemer \& George 2003).  Observations in
the ultraviolet (Crenshaw \et 1999; Crenshaw \& Kraemer 2001) showed
\ark\ to be one of the Seyfert 1 galaxies that showed no intrinsic
ultraviolet absorption. Thus \ark\ is a rare example of a bright
Seyfert 1 galaxy that is not significantly affected by any kind of
complex absorption -- its emission spectrum is that of a `bare'
Seyfert 1 nucleus.

The paper presents the results of a long \xmm\ observation designed to
characterise the intrinsic X-ray emission spectrum of a Seyfert 1
galaxy. The rest of this paper is organised as follows.
Section~\ref{sect:obs} discusses the observation details and data
reduction. Section~\ref{sect:timing} gives details of the X-ray
variability observed in \ark\ and  section~\ref{sect:fluxed} 
describes a `first look' at the X-ray spectrum.
The spectrum is then examined in detail using the Reflection Grating
Spectrometer (RGS) in section~\ref{sect:rgs}. This is followed by
an analysis of the European Photon Imaging Camera (EPIC) data
first over the $3-10$~keV band (section~\ref{sect:epic}) and then
the over the full band-pass (section~\ref{sect:broad}).
Section~\ref{sect:xte}  briefly discusses an
analysis of archival \xte\ observations of \ark.  Finally, the results
are discussed in section~\ref{sect:disco}.

% --------------------------------------------------------------------------
% --------------------------------------------------------------------------
% --------------------------------------------------------------------------

\section{Observation and data reduction}
\label{sect:obs}

\xmm\ (Jansen \et 2001) observed \ark\ during the period 2003 August
24 05:35:43 -- 2003 August 25 12:44:33.  \xmm\ carries three
co-aligned X-ray telescopes each with an
EPIC CCD array as its focal-plane detector.  There are two types of
EPICs used: two MOS cameras (Turner \et 2001a) and one pn camera
(Str\"{u}der \et 2001). The two telescopes that focus onto the MOS cameras
also focus onto RGS instruments
(den Herder \et 2001).  During the \ark\ observation all the EPIC
cameras were operated in small window mode using the thin
filter.  The short frame-time possible in small window mode meant
photon pile-up (which can distort the spectrum of bright sources;
Ballet 1999) was negligible.

The extraction of science products from the Observation Data Files
(ODFs) followed standard procedures using the \xmm\ Science Analysis
System ({\tt SAS v5.4.1}).  The EPIC and RGS data were processed using
the standard {\tt SAS} processing chains to produce calibrated event
lists.  These removed events from the position of known defective
pixels, corrected for Charge Transfer Inefficiency (CTI) and applied a
gain calibration to produce a photon energy for each event.  Source
data were extracted from circular regions of radius $45$\arcs from the
processed images and background events were extracted from regions in
the small window least effected by source photons.  These showed the
background to be relatively low and stable throughout the observation.
The total amount of `good' exposure time selected was $78.2$~ks from
the pn, $108.4$~ks from the MOS and $111.7$~ks from the RGS.  (The
lower pn exposure is due to 
the lower `live time' of the pn camera in small-window mode, $\sim 71$
per cent; Str\"{u}der \et 2001).  The total number of source photons
extracted was $2.2 \times 10^6$ for the pn, $8.0 \times 10^5$ for each
MOS, and $8.6 \times 10^4$ and $9.9 \times 10^4$ counts for the RGS1
and RGS2, respectively. 

The spectral analysis was performed using {\tt XSPEC v11.2} (Arnaud
1996) and, in order to allow effective use of the method of $\chi^2$
minimization for spectral fitting, the EPIC spectra were grouped such
that each energy bin contains at least $50$ counts. The RGS data were
grouped to contain at least $20$ counts per bin.  The quoted errors on
the derived model parameters correspond to a $90$ per cent confidence
level for one interesting parameter (i.e. a $\Delta \chi^{2}=2.71$
criterion), unless otherwise stated, and fit parameters are quoted for
the rest frame of the source.   Values of $ H_{0} = 70 $~km s$^{-1}$
Mpc$^{-1}$ and $ q_0 = 0.5 $ are assumed throughout the paper.

% --------------------------------------------------------------------------
% --------------------------------------------------------------------------
% --------------------------------------------------------------------------

\section{Timing analysis}
\label{sect:timing}

\begin{figure}
\centering
\includegraphics[width=6.40 cm, angle=270]{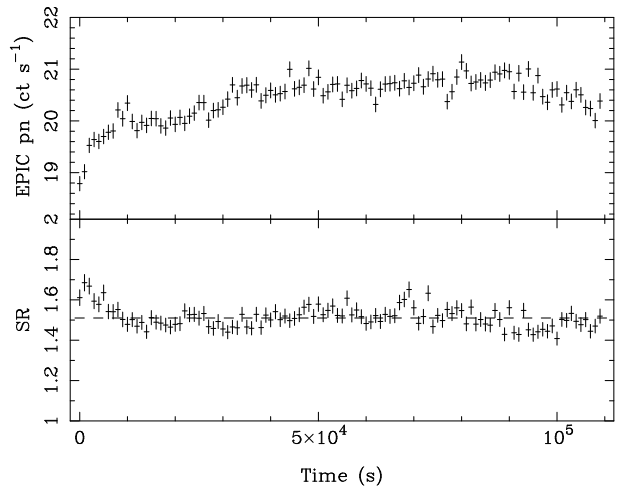}
\caption{
Top panel: 
EPIC pn light curve of \ark\ ($0.2-10$~keV with $1000$~s bins).
Note the false zero on the ordinate.
The source clearly shows significant but low amplitude
flux variations.
Bottom panel: softness ratio ($0.2-0.5/2-10$~keV) showing
weak but significant spectral variations.
}
\label{fig:lc}
\end{figure}

Fig~\ref{fig:lc} shows the broad band X-ray light curve obtained
using the pn camera.  This light curve has been background subtracted
and corrected for exposure losses.  The source shows only very mild
variations in flux. Fitting the light curve with a constant  gave an
unacceptable fit ($\chi^2 = 929.8$ for $109$ degrees of freedom),
rejecting the constant hypothesis with very high significance
($>99.99$ per cent) and demonstrating the observed variations are
intrinsic to the source.  The measured amplitude of
variations,  estimated using the $F_{\rm var}$ statistic (see Vaughan
\et 2003), was only $F_{\rm var} = 1.9$ per cent.

\begin{figure}
\centering
\includegraphics[width=6.40 cm, angle=270]{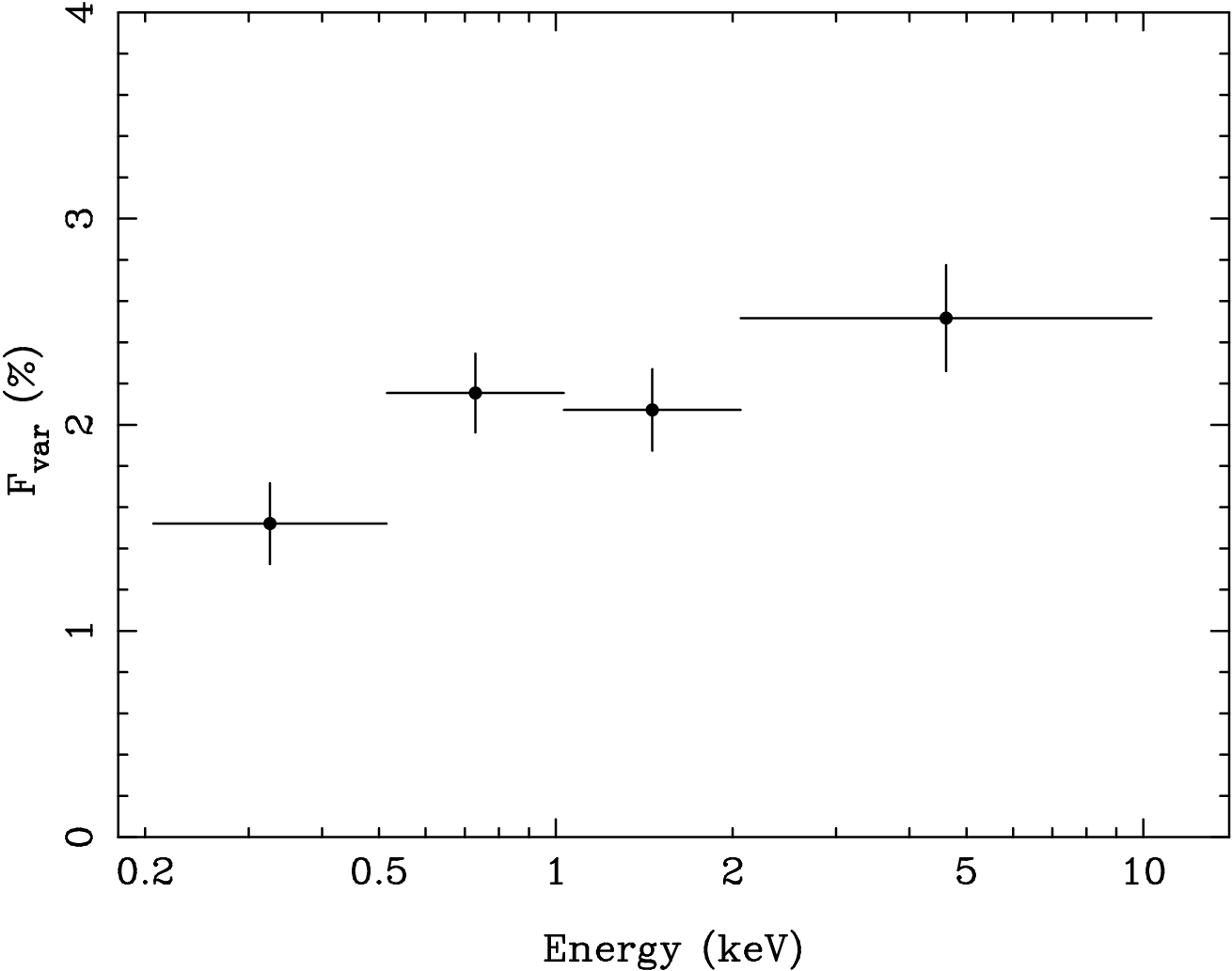}
\caption{
Rms spectrum of \ark.
}
\label{fig:rms}
\end{figure}

The data were divided into four energy bands ($0.2-0.5$, $0.5-1.0$,
$1-2$ and $2-10$~keV) in order to investigate the spectral dependence
of the variations. 
The individual light curves show similar trends.
The ratios of these light curves in the four bands
were used to test of spectral variability. A constant was fitted to
each of the ratios and the resulting $\chi^2$ value recorded.
Spectral variability was clearly detected, with a constant hypothesis
being rejected at $>90$ per cent confidence in five of the six
ratios, and rejected at $>99.99$ per cent confidence for the
ratio $0.2-0.5/2-10$ ($\chi^{2} = 233.119 / 109~ dof$). 
The time series of this ratio is shown in Fig.~\ref{fig:lc}. 
The variability amplitude was measured in each of the four
bands and used to define the rms spectrum (Fig.~\ref{fig:rms}). 
There was an obvious enhancement of variability amplitude at higher
energies. 

% --------------------------------------------------------------------------
% --------------------------------------------------------------------------
% --------------------------------------------------------------------------

\begin{figure}
\centering
\includegraphics[width=6.40 cm, angle=270]{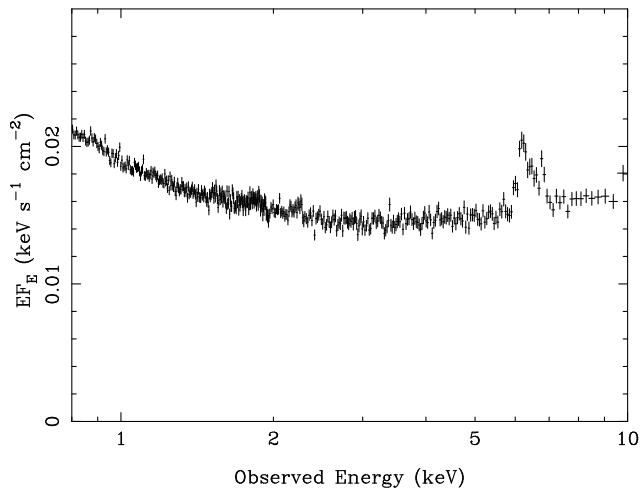}
\caption{
The `fluxed' spectrum of \ark\
after  correcting for Galactic absorption.
This was produced by calculating the ratio of 
the spectra of \ark\ and \3c\ and then normalising by the spectral
model for \3c (in flux units). }
\label{fig:spectrum}
\end{figure}

\section{A first look at the X-ray spectrum}
\label{sect:fluxed}

The X-ray spectrum in true flux units was estimated using the
procedure discussed in Vaughan \& Fabian (2004).  The counts to flux
conversion was carried out using an \xmm\ observation of \3c\ as a
`standard star.'  \3c\ is a bright source and has a relatively simple
spectrum in the EPIC band (a hard power-law plus smooth soft
excess modified by Galactic absorption; Page \et 2004a), and in
particular does not contain any strong, sharp spectral features.
The ratio of the spectra of \ark\ and \3c\ was calculated
and then normalised by a
spectral model for \3c\ (see Vaughan \& Fabian 2004 and also Page \et
2004a). Fig~\ref{fig:spectrum} shows the resulting spectrum and
clearly reveals the strongest spectral features in \ark, namely the
smooth, featureless continuum and the iron line peaking at $\approx
6.4 \keV$. 

This spectrum was corrected for Galactic absorption.  The low Galactic
latitude of \ark\ ($b=-21.1\deg$) means foreground Galactic absorption
is high.  The column density towards \ark\ has been estimated from
$21$~cm surveys of the surrounding sky to be  $N_{\rm H} = 0.98 \times
10^{21}$~cm$^{-2}$  (using the data of Stark \et 1992) or $N_{\rm H} =
1.26 \times 10^{21}$~cm$^{-2}$ (using the data of Dickey \& Lockman
1990).  Either of these values can be consistent with the estimate
based on fits to the \rosat\ soft X-ray spectrum (Brandt \et 1993),
although which one depends on the assumed form of the underlying
spectrum. In the analysis below the  Galactic absorption was assumed
to be given by the smaller of these two column density estimates
(unless stated otherwise) and was modelled using the {\tt TBabs} code
of Wilms, Allen \& McCray (2001).  The relatively large correction for
\ark\ introduced spurious features into the fluxed spectrum at low
energies.  This is because the source spectrum was convolved through
the detector response but the absorption model used for the correction
was not. This lead to a spurious `inverted edge' feature appearing
near the Galactic O~\textsc{i} edge.   Furthermore, the \3c\
observation used the medium filter while the \ark\ observation used
the thin filter, this will systematically distort the fluxed spectrum
of \ark\ at the lower energies.  Due to these issues the fluxed
spectrum is shown only above $0.8$~keV.

\begin{figure}
\centering
\includegraphics[width=7 cm, angle=270]{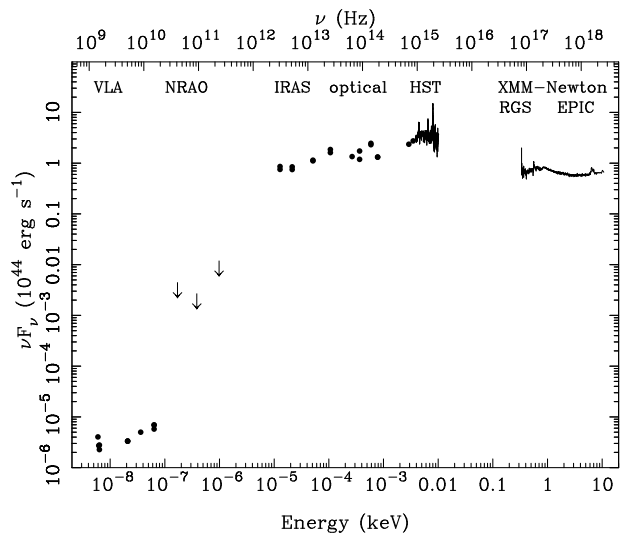}
\caption{
The `fluxed' X-ray spectrum of \ark\ (see Figs.~\ref{fig:spectrum} and
\ref{fig:rgs_1}) and the non-simultaneous, multi-wavelength SED.
}
\label{fig:sed}
\end{figure}

Figure~\ref{fig:sed} shows the X-ray spectrum compared to the overall
SED obtained from non-simultaneous observations.  The multi-wavelength
fluxes (VLA, NRAO, IRAS, optical) were obtained from the  NASA/IPAC
Extragalactic Database (NED).  The ultraviolet \hst\ spectrum was
taken with the  Faint Object Spectrograph (FOS) on-board \hst\ and was
previously published by Crenshaw \et (1999).  These data were obtained
through the Multi-Mission Archive at Space Telescope (MAST) and
de-reddened assuming $E(B-V)=0.128$.   The SED was produced using data
obtained over a number of years by a number of observers and so should
not be taken to represent the single-epoch SED of \ark.  Nevertheless
this SED does provide a strong indication of the general shape of the
broad-band SED of \ark, which must peak in the far-ultraviolet region.
Interestingly, the soft X-ray excess (see section~\ref{sect:broad}) does
not appear steep enough to be the high energy tail of the
blue-ultraviolet bump.

% --------------------------------------------------------------------------
% --------------------------------------------------------------------------
% --------------------------------------------------------------------------

\section{RGS spectral analysis}
\label{sect:rgs}

\begin{figure}
\centering
\includegraphics[width=6.40 cm, angle=270]{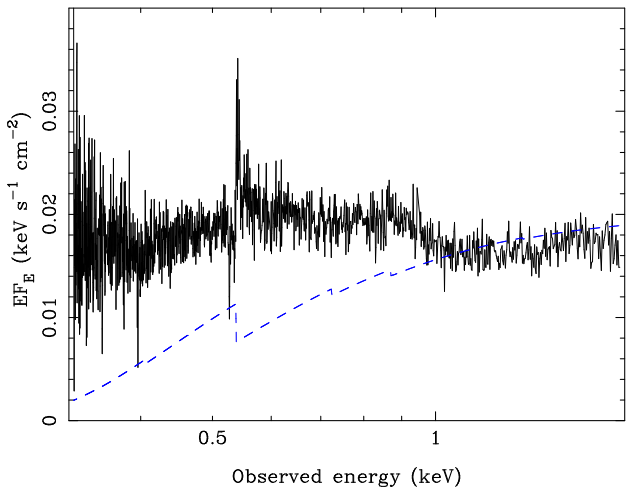}
\caption{
Combined RGS spectrum shown in flux units after
removing the Galactic absorption.
(The dashed line shows the Galactic absorption profile 
scaled by a factor $2\times 10^{-2}$.)
Note that the emission line-like feature at $\sim 0.54$~keV 
is largely an artifact of the absorption correction
(compare with Fig~\ref{fig:rgs_2}).
The nearby absorption line is real and due to the interstellar 
O~\textsc{i} $1s-2p$ line. 
For clarity the error bars are not shown.
}
\label{fig:rgs_1}
\end{figure}

\begin{figure*}
\centering
\includegraphics[width=12 cm, angle=270]{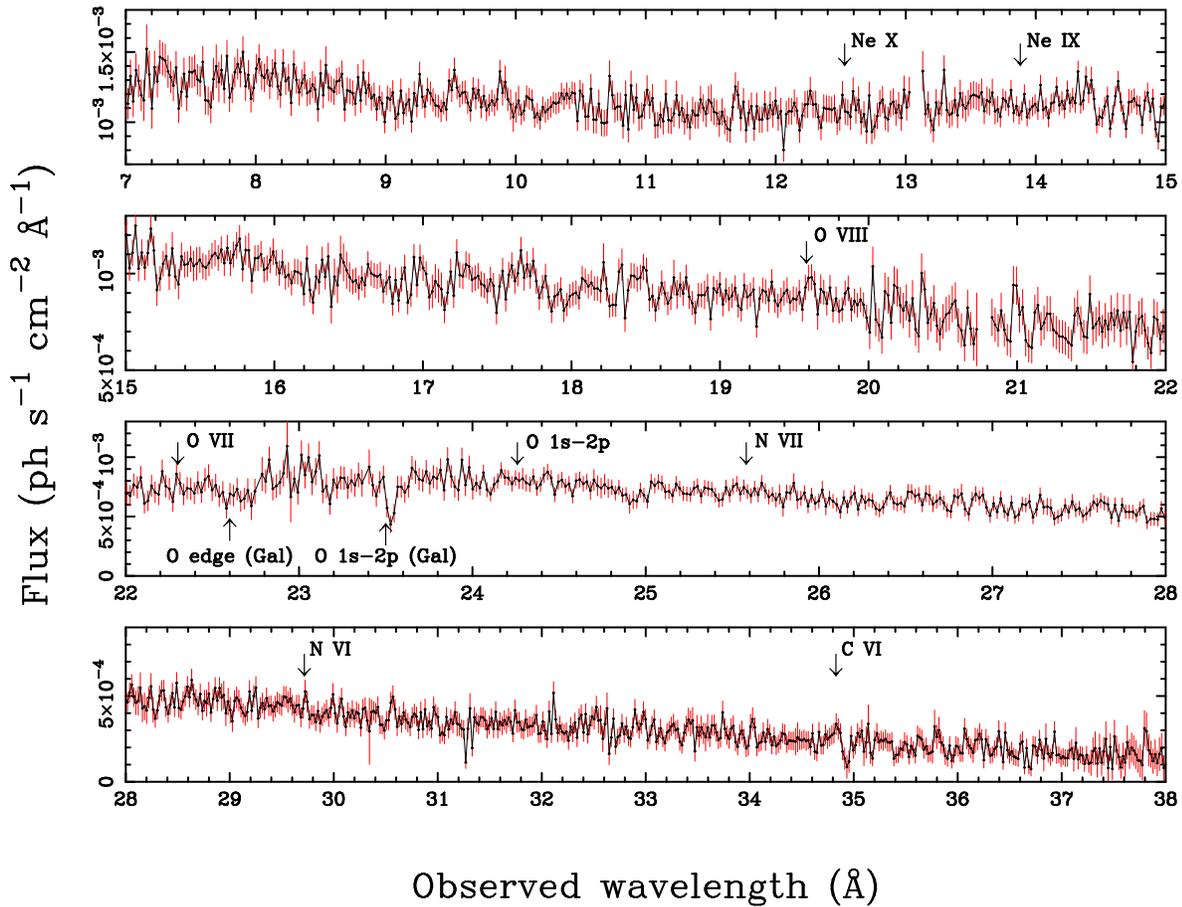}
\caption{
Close-up of the fluxed RGS spectrum plotted in wavelength units.
The expected wavelengths of Galactic absorption features 
(the K-edge and principal resonance line of
interstellar O~\textsc{i}) are indicated as upwards arrows. 
Clearly the Galactic O~\textsc{i} resonance line is detected.
In addition, the wavelengths (shifted by the systemic velocity of
\ark) of the principal resonance lines of highly ionized C, N, O and
Ne, as well as neutral O, are shown with downwards
arrows (all undetected).
The slight `dip' longwards of C~\textsc{vi} coincides with
a bad column in the RGS2 spectrum and is unlikely to be intrinsic to \ark.
}
\label{fig:rgs_2}
\end{figure*}

The combined RGS1+RGS2 spectrum is shown in true flux units (as a
function of energy) in Fig~\ref{fig:rgs_1}.
This was corrected for Galactic absorption using the {\tt TBabs}
model (which accounts only for photoelectric absorption) and assuming
$N_{\rm H} = 0.98 \times 
10^{21}$~cm$^{-2}$. Due to small inaccuracies in both the
RGS response and the interstellar absorption model (see de Vries \et
2003) this resulted in a spurious `spike' in the corrected data at
$\sim 0.54$~keV.
Fig~\ref{fig:rgs_2} shows a closer examination of the fluxed RGS
spectrum as a function of wavelength.  
In this case no correction was made for the Galactic absorption and
the only obvious features were those
due to absorption by interstellar O.  Excepting this, the
intrinsic soft X-ray spectrum appears, to first order at least, as a
smooth continuum with no obvious, strong, discrete features.

\subsubsection{Intrinsic warm absorption}

In order to test for traces of ionized (warm) absorption local to
\ark\ the RGS data were searched for resonance absorption lines of C,
N, O and Ne. In particular, the (source frame) wavelengths
corresponding to the $1s-2p$ resonance transitions in He- and H-like
ions of N, O and Ne as well as H-like C were examined. 
The relevant atomic data (wavelengths and oscillator strengths) were
obtained from Verner, Verner \& Ferland (1996).
The expected (source frame) wavelengths of these transitions are shown
in Fig.~\ref{fig:rgs_2}. 

In order to measure the parameters of each line
a region of the spectrum $\sim 2$~\A\ either side of the expected
transition was fitted with a 
power-law modified by Galactic absorption. A narrow ($\sigma = 0.1$~eV)
absorption line was then added to the data at the wavelength expected
for the transition (after correcting for the redshift).
In the case of all seven possible transitions no significant improvement in
the fit was found upon adding an absorption line to the model.
Upper limits were obtained on the equivalent widths of these lines
as shown in Table~\ref{tab:abs}. 
These limits were derived after allowing for moderate
red/blue shifts (up to $7000$~km s$^{-1}$) in the lines.
Changing the wavelength range fitted did not
substantially alter these results.

These equivalent width limits place constraints on the possible
columns of ionized absorbing material as follows.
Any weak absorption lines are likely to be unsaturated and
thus on the linear part of the curve-of-growth.
For a given transition the measured equivalent width thus  
corresponds to an ionic column density given by the following (Spitzer
1978): 

\begin{equation}
\frac{EW_{\lambda}}{\lambda} =
\frac{\pi {\rm e}^2}{m_{\rm e} c^2} N_j \lambda f_{ij} =
8.85 \times 10^{-13} N_j \lambda f_{ij}
\end{equation}

where $EW_{\lambda}$ is the equivalent width and $\lambda$ is the
wavelength (in cm) of the line, $N_j$ is the ionic column density
and $f_{ij}$ is the oscillator strength.
The derived limits on the ionic column densities are given in
Table~\ref{tab:abs}. 

\subsubsection{Interstellar absorption}

As mentioned above, the column density of Galactic interstellar gas is
uncertain but estimated to lie in the range $\sim 1.0-1.3 \times
10^{21}$ cm$^{-2}$. The RGS spectrum clearly shows discrete structure
due to absorption by local interstellar O~\textsc{i} (see
Fig.~\ref{fig:rgs_2}; note that the region of the spectrum around
O~\textsc{i} is sampled only by RGS1).  
The depth of the O~\textsc{i} $1s-2p$
absorption line can provide an independent estimate of the column
density. The line is a simpler diagnostic than the O~\textsc{i} K-edge
which is broader and complicated by instrumental absorption (see de
Vries \et 2003). The equivalent width of the line ($EW \sim 60$~m\A)
corresponds to an absorbing column density 
$\log(N_{\rm O}) \sim 17.6-18.0$ (see Fig 3 of de Vries 2003).
The
main source of uncertainty  in the column density estimate is the
turbulent velocity width of the line; the line is deep enough that it
does not lie on the linear part of the curve-of-growth. 
This column of neutral O corresponds to an equivalent hydrogen
column density of $N_{\rm H} 
\approx 0.8-2.0 \times 10^{21}$~cm$^{-2}$ (assuming an O/H abundance of
$4.90 \times 10^{-4}$ as given by Wilms \et 2000). 
The column density estimate is therefore consistent with the $21$~cm
measurements but does not improve upon them.

The neutral absorption intrinsic to the host galaxy of \ark\  was
estimated by measuring the O~\textsc{i} $1s-2p$ line in
the redshifted frame
of the galaxy. The limit on its equivalent width is given in
Table~\ref{tab:abs}. This corresponds to column densities of
$N_{\rm O} \ls 10^{17}$ cm$^{-2}$ and 
$N_{\rm H} \ls 2 \times 10^{20}$ cm$^{-2}$. The interstellar
absorption intrinsic to \ark\ is therefore negligible compared to the
foreground absorption.

\begin{table}
\centering
\caption{
Equivalent width estimates for $1s-2p$ absorption lines as derived from the RGS
data. The wavelengths are the expected rest-frame wavelengths from Verner \et (1996).
$^{a}$ Galactic absorption line (measured at $z=0.0$).
$^{b}$ absorption line intrinsic to \ark\ (measured at $z=0.0323$).
\label{tab:abs}
}
\begin{tabular}{lrrr}
\hline
line   & $\lambda$~(\A) & $EW$~(m\A) & $N_{i}~({\rm cm}^2)$\\
\hline
C~\textsc{vi}      & $33.736$        & $<64$    & $<1.5\times10^{16}$\\
N~\textsc{vi}      & $28.787$        & $<30$    & $<6.1\times10^{15}$\\
N~\textsc{vii}     & $24.781$        & $<12$    & $<5.3\times10^{15}$\\
O~\textsc{vii}     & $21.602$        & $<28$    & $<9.7\times10^{15}$\\
O~\textsc{viii}    & $18.969$        & $<10$    & $<7.5\times10^{15}$\\
Ne~\textsc{ix}     & $13.447$        & $<2$     & $<1.7\times10^{15}$\\
Ne~\textsc{x}      & $12.134$        & $<12$    & $<2.2\times10^{16}$\\
\hline
O~\textsc{i}$^{a}$ & $23.50$        & $60\pm13$ & $6-10\times10^{17}$\\
O~\textsc{i}$^{b}$ & $23.50$        & $<20$     & $<1\times10^{17}$\\
\hline
\end{tabular}
\end{table}

% --------------------------------------------------------------------------
% --------------------------------------------------------------------------
% --------------------------------------------------------------------------

\section{EPIC $3-10$~keV spectral analysis}
\label{sect:epic}

The fluxed spectrum (section~\ref{sect:fluxed}) demonstrated 
that the X-ray emission from  \ark\
appears as a reasonably smooth continuum and the RGS 
showed no evidence for warm absorption (section~\ref{sect:rgs}). 
This section describe an analysis of the 
EPIC spectrum, starting with an examination of the
iron line.

\subsection{Iron K$\alpha$ emission line}
\label{sect:iron}

The $3.5-10$~keV spectrum was fitted, after excluding the $5-8$~keV
region which may contain Fe K features, with a  simple model
comprising a power-law modified only by Galactic absorption.  The pn
and MOS spectra gave differing spectral slopes.  The photon index of
the power-law fitted to the pn data was $\Delta \Gamma \approx 0.1$
steeper than that fitted to the MOS data.  The two MOS spectra were in
good agreement with each other.   Similar discrepancies have been seen
in other observations (Molendi \& Sembay 2003; Vaughan \& Fabian
2004).  Hence, in the following spectral analysis, the power-law slope
was always  allowed to differ between the pn and the MOS (but with the
slope for the two MOS spectra tied together).

\begin{figure}
\centering
\includegraphics[width=5.60 cm, angle=270]{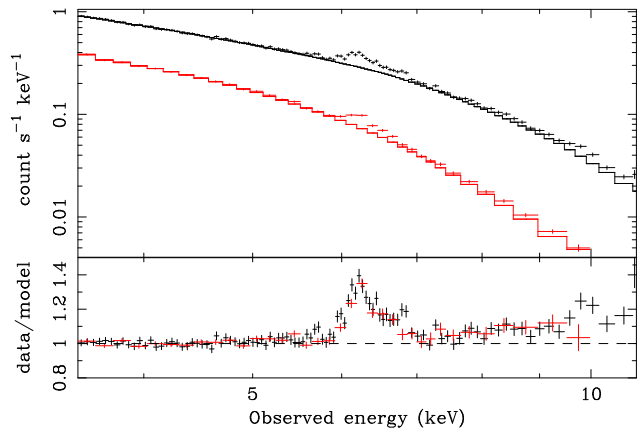}
\caption{
EPIC pn and combined MOS spectrum of \ark.
The data were fitted with a power-law
over only the $3.5-5$~keV `continuum' region. The model was then
extrapolated up in energy to reveal the excess 
due to the Fe emission.
(NB: the MOS1 and MOS2 data were fitted separately
and combined only for plotting purposes.)
}
\label{fig:line}
\end{figure}

The fluxed spectrum clearly showed Fe emission (Fig~\ref{fig:spectrum}),
if this is produced by fluorescence in an optically thick medium
then there may also be an associated Compton reflection continuum (Lightman \&
White 1988; George \& Fabian 1991), which would show as a flattening
of the spectrum at higher energies.
Thus the spectral slope was estimated using only the $3.5-5.0$~keV
region -- this is free from absorption effects and
contains no significant contribution from either the line or
reflected flux.
The slope was measured to be $\Gamma = 2.01\pm0.06$ (from the pn).
Fig~\ref{fig:line} shows the residuals from this fit after the
model was extrapolated to high energies.  The excess
residuals due to Fe emission are obvious. 
There also appears to be a slight up turn in the continuum at
energies above the line which could be due to reflection (see
sections~\ref{sect:refln} and \ref{sect:xte}). 

The line could in
principle be composed of  lines at energies of $6.4-6.9$~keV which
could have profiles ranging from unresolved ($\sigma \ls 70$~eV) to
very broad and asymmetric.  A set of trial models were compared to the
data to test the most plausible possibilities for the line structure.
The continuum was assumed to be a simple power-law between
$3.5-10$~keV (modified by Galactic absorption). 
The line was then fitted with a variety of alternative models
including unresolved, narrow lines (modelled using narrow
Gaussians of width $\sigma = 10$~eV), broad Gaussians and
relativistically broadened line profiles (the {\tt diskline} model of
Fabian \et 1989).

\begin{table*}
 \caption{Best-fitting parameters for models of the iron line.
Column (1) describes the line model components, either
broad/narrow Gaussians or diskline profiles.
Columns (2) and (3) give the line energies and equivalent widths.
Column (4) gives the Gaussian width ($\sigma$) or emissivity index
($q$) in the case of diskline models.
Columns (5) and (6) give the inner and outer radii and
column (7) gives the inclination angle for diskline components.
Column (8) lists the fits statistics.
$^{f}$ indicates the parameter was fixed.
}
\centering
\label{tab:line}            
\begin{tabular}{llrrrrrrr}
\hline
   &            & E             & EW            & $\sigma/q$ & \rin         & \rout         & $i$           &               \\
   & Model       & (keV)         & (eV)          & (eV)          & (\rg)         & (\rg)         & (deg)         & $\chi^{2}/dof$\\
   & (1)         & (2)           & (3)           & (4)           & (5)           & (6)           & (7)           & (8)           \\
\hline
$1$  & narrow      & $6.40\pm0.01$ & $70\pm6$      & $10^f$        &               &               &               & $1717.2/1520$ \\ 
\\
$2$  & narrow+     & $6.40\pm0.01$ & $66\pm5$      & $10^f$        &               &               &               & $1633.0/1518$ \\ 
     & narrow      & $6.66_{-0.05}^{+0.01}$ & $28\pm5$      & $10^f$ &             &               &               &               \\
\\
$3$  & narrow+     & $6.34\pm0.02$ & $44\pm9$      & $10^f$        &               &               &               & $1580.0/1516$ \\ 
     & narrow+     & $6.50_{-0.04}^{+0.01}$ & $41\pm7$      & $10^f$ &             &               &               &               \\
     & narrow      & $6.76\pm0.02$ & $26\pm5$      & $10^f$        &               &               &               &               \\
  \\ 
$4$  & broad       & $6.45\pm0.02$ & $130\pm14$    & $177_{-26}^{+31}$ &           &               &               & $1596.0/1519$ \\ 
\\
$5$  & broad+      & $6.57\pm0.05$ & $109\pm16$    & $288_{-38}^{+41}$ &           &               &               & $1545.1/1517$ \\ 
     & narrow      & $6.40\pm0.01$ & $37\pm7$      & $10^f$        &               &               &               &               \\
\\
$6$  & diskline    & $6.42\pm0.01$ & $156\pm15$    & $-2.24\pm0.12$ & $57_{-39}^{+27}$ & $>10^5$ & $>48$            & $1566.3/1516$ \\ 
\\
$7$  & diskline+   & $6.56_{-0.06}^{+0.02}$ & $99_{-6}^{+11}$      &  $-2.81_{-0.19}^{+0.23}$ & $144_{-30}^{34}$ & $>10^5$ & $>73$ &  $1541.8/1514$ \\ 
     & narrow      & $6.40\pm0.01$ & $42\pm6$      & $10^f$        &               &               &               &               \\
\hline
\end{tabular}
\end{table*}

The best-fitting parameters and fit statistics are listed in
Table~\ref{tab:line}. The seven different models tested were as
follows:  
($1$) one, ($2$) two and ($3$) three narrow Gaussians;  
($4$) one broad Gaussian; 
($5$) one broad and one narrow Gaussian;  
($6$) a diskline;  
($7$) a diskline and one narrow Gaussian.  
The data were not consistent with a model for the
iron line comprising just one or two unresolved emission lines.  That
said, the data did not strongly favour any one of the more complex models,
models $3-7$ all provided reasonable fits to the data (rejection
probability $<0.95$).  However, the two models that include both a
broad and a narrow line (models $5$ and $7$) gave superior
fits.

\begin{figure}
\centering
\includegraphics[width=5.60 cm, angle=270]{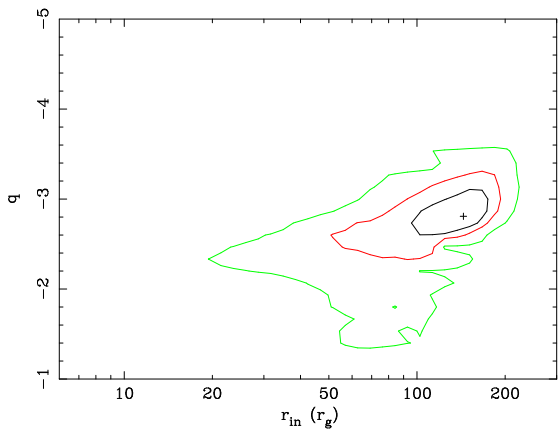}
\caption{
Contours of $\Delta \chi^2$ around the best-fitting
diskline model (section~\ref{sect:iron}) as a function of
disc emissivity index $q$ and inner radius $r_{\rm in}$.
The contours delineate the $\Delta \chi^2=2.3, 4.61, 9.21$ regions.
The best fitting solution (marked with a cross) 
has a large inner radius
but the $\Delta \chi^2$-space is complicated and
lower inner radius solutions can be allowed provided
the emissivity law is flatter.
The other diskline parameters (rest-frame line energy,
outer radius and inclination angle) were free parameters.
}
\label{fig:contour}
\end{figure}

The best description for the iron line in \ark\ is thus in terms of a
narrow line (with equivalent width $\approx 40$~eV) and a stronger,
broader line (with equivalent width $\approx 100$~eV).  The broad line
was described well in terms of either a Gaussian or a diskline
profile. In the case of the broad Gaussian, the line width ($\sigma
\approx 290$~eV) corresponds to a velocity width of $FWHM \approx 3
\times 10^4$~km s$^{-1}$. This is far broader than the broad,
permitted optical lines ($FWHM({\rm H}\beta) \approx 5800$~km
s$^{-1}$; Wandel \et 1999) and so places the X-ray line emitting
region within the optical Broad Line Region (BLR).  Assuming the line
originates from a Keplerian disk about a black hole the width measured
from the broad Gaussian fit places the line emitting material within a
few hundred \rg\ of the black hole (depending on the inclination).  

In the case of the diskline model the line profile was broad but
reasonably symmetric -- the large inner disk edge  and high
inclination angle mean Doppler shifts will dominate over gravitational
redshift.  The inner disc radius was constrained to lie  beyond
$r_{\rm in}=6\rg$, the  innermost stable circular orbit (ISCO)
expected for an accretion disc about a non-spinning (Schwarzschild)
black hole. Fixing the inner radius to $r_{\rm in}=6\rg$ (in model
$7$) gave a significantly worse fit even with all the other line
profile parameters left free ($\Delta \chi^2 = 14.0$).  That said, the
$\chi^2$-space is complicated and smaller inner radii can be permitted
provided the emissivity index ($q$ where emissivity is $J(r)\propto
r^{q}$) is flatter and inclination angle is lower,
as illustrated by Fig.~\ref{fig:contour}. 
Nevertheless $r_{\rm
in}=6\rg$ always lies outside the $99$ per cent  confidence region of the
best-fitting model (model $7$).

The best-fitting energy of the line is rather unusual, at $E \approx
6.56$~keV in both Gaussian and diskline models.  A line energy of
$6.4$~keV, corresponding to neutral iron, gives a substantially worse
fit to the data ($\Delta \chi^2=21.5$ compared to model 7,
Table~\ref{tab:line}). This best-fitting energy corresponds to the
line emission being dominated by mildly ionized iron
(Fe~\textsc{xx-xxii}) meaning that resonant trapping and the Auger
effect (see e.g. Ross, Fabian \& Brandt 1996) should destroy the line.

\subsubsection{Additional iron line complexity}

The high inner radius of the {\tt diskline} model can be alleviated if
the intrinsically narrow line component is replaced by a second {\tt
diskline}. Fixing $r_{\rm in}=6$ and $r_{\rm out}=10^{4} \rg$ for the
pair of lines (with energies and emissivity indices free to vary
independently) gave a good fit ($\chi^{2} = 1545.8 / 1515 ~ dof$),
comparable to the best fits given in table~\ref{tab:line}. The line
energies and emissivity indices were as follows: $E_1 = 6.41\pm
0.01$~keV, $q_1 = -1.8_{+0.2}^{-0.1}$ and $E_2 = 6.80_{-0.40}^{+0.04}$~keV,
$\beta_2 = -2.2_{+1.0}^{-0.2}$.   The best-fitting inclination angle
was $i = 32\pm5\deg $.  Thus it remains plausible that $r_{\rm in}=6$
as long as there  are two components to the Fe K$\alpha$ line  from
the disc, each with flat emissivity laws.

The only noticeable residuals that remained after the
narrow and broad lines had been fitted (with either model $5$ or $7$),
appeared as a slight excess 
at $E\approx 7$~keV and a slightly deficit at $E \approx 6$~keV. 
The former of these could plausibly be accounted for by emission
at $6.9$~keV by the Ly$\alpha$ line of H-like Fe, or $7.06$~keV
emission the by the K$\beta$ line of neutral Fe, or some combination of the
two.  Adding an additional narrow Gaussian to model 5
(Table~\ref{tab:line})  improved the fit by $\Delta \chi^2 = 9.0$ with
best fitting energy $E=7.02 \pm 0.05$~keV and an equivalent width $EW
= 15 \pm 8$~eV.  The intrinsic width was poorly constrained when left
free ($\sigma < 246$~eV). The ratio of flux in the main Fe lines to
the flux in the $\approx 7$~keV line was $\approx 7.8$ for the broad
line and $\approx 2.6$ for the narrow line. The expected ratio of
K$\alpha$/K$\beta$ flux is $150/17 \approx 8.8$ for neutral iron. Thus the
weak line at  $\approx 7$~keV could be accounted for as the K$\beta$
line accompanying the broader of the two K$\alpha$ lines. However,
some contribution from Fe Ly$\alpha$ and/or a narrow, neutral K$\beta$
line cannot be ruled out. 

The negative residual at $E \approx 6$~keV again seemed marginally
significant. Including a narrow Gaussian absorption line improved the
fit by a further $\Delta \chi^2 = 9.0$ with best fitting energy
$E=6.03_{-0.04}^{+0.01}$~keV and an equivalent width $EW = -11 \pm
6$~eV. The identification of this feature is less obvious. As it is
only a tentative detection, rather weak (in equivalent width terms)
and has no clear identification, a detailed analysis of this possible
feature is not given in this paper.

\subsection{Reflection spectrum}
\label{sect:refln}

As noted above the X-ray continuum appears to be slightly curved
compared to a power-law. This curvature is in the same sense as
expected if there is a Compton reflection continuum component present
in the data. Furthermore, an archival \xte\ spectrum covering the
$3-20$~keV range (see section~\ref{sect:xte}) supports the presence of
a reflection continuum. A simple hypothesis would be that the spectral
curvature in the $3.5-10$~keV EPIC spectrum is caused by  a reflection
continuum which originates in the same material as that responsible
for the iron line.  Assuming a neutral slab with solar abundances
(George \& Fabian 1991; Matt, Fabian \& Reynolds 1997)
the equivalent width of the broad line ($EW \approx 100$~eV)
corresponds to a reflection strength of $R \approx 0.5 - 1$
(where $R$ represents the ratio of the normalizations of the observed
and reflected power-law continua) depending on the viewing inclination.

As a first test of this, a reflection continuum was included in the
model  using the {\tt pexrav} code (Magdziarz \& Zdziarski 1995).
This component, which accounts for only the continuum produced by
Compton reflection and photoelectric absorption
(not any associated fluorescent/recombination emission) 
was added to the basic model of a power-law continuum
plus two Gaussian iron lines (model $6$ of Table~\ref{tab:line}).
The quality of the fit was only very slightly improved ($\chi^{2} =
1542.2/ 1516~dof$) and the relative reflection strength  was
constrained to lie in the range $R=0.02 - 0.89$. This calculation was
performed assuming the reflector was inclined at $i = 30 \deg$ with
respect to the line of sight but almost identical constraints were
obtained assuming an inclination angle of $i = 60 \deg$. Thus the
\xmm/EPIC data are consistent with the presence of a moderate strength
Compton reflection component ($R \ls 1$), which would match the
strength of the iron line, but are not sufficient to give a
significant detection (but see section~\ref{sect:xte}).
The line parameters do not change substantially upon the inclusion of
a reflection component to the model.

In order to test for the possible presence of a strongly
gravitationally redshifted emission line this reflection plus double
Gaussian model  was refitted including an addition {\tt Laor} emission
profile  (Laor 1991). The inner and outer radii of the disc were fixed
at $r_{\rm in}=1.24$ and $r_{\rm out}=20$ to simulate only the very
innermost region of a disc extending close to the black hole. Assuming
a neutral line and an
emissivity power-law index of $q=-3$, the upper limit on
the equivalent width of this feature was $EW \ls 7, 21, 131 $~eV
for inclination angles of $i=30\deg, 60\deg, 80\deg$
respectively.  At high inclinations, the line emission from this
region of the disc is smeared over  such a large energy range that,
even in this relatively high quality spectrum, it was difficult to
constrain. It is clear however that unless the disc inclination
angle is high $ i > 60\deg $ there can be no strong, gravitationally
redshifted component to the line.

The possibility of reflection 
was examined further using the reflection code of
Ross \& Fabian (1993) to model the line emission and
Compton reflection continuum self-consistently. The computed
reflection spectrum was convolved with a {\tt diskline}
kernel to account for possible Doppler and gravitational broadening.
The ionization of the reflector was kept fixed at 
its minimum allowed value ($\xi = 4 \pi F_{\mathrm{X}}/n_{\mathrm{H}}
= 10$) and an additional narrow, neutral ($6.4$~keV) iron line was
included using a ($\sigma=10$~eV) Gaussian profile. This model therefore accounts for
both the strength and profile of the broad iron line as well as the
strength of the Compton reflection continuum. 
The best-fitting model provided an acceptable fit to the data
($\chi^2 = 1552.7 / 1518 ~ dof$) with parameters as follows: 
$\Gamma = 1.96 \pm 0.02$, 
$R=0.49\pm0.07$, 
$\rin = 212_{-133}^{+109} \rg$ and
$i \ge 43 \deg$.
These results are consistent with both the {\tt diskline} model discussed
above (section~\ref{sect:iron}) and the \xte\ fitting described in
section~\ref{sect:xte}, and show the 
broad iron line and weak reflection continuum are consistent
with a common origin.

% --------------------------------------------------------------------------
% --------------------------------------------------------------------------
% --------------------------------------------------------------------------

\section{The broad-band EPIC spectrum}
\label{sect:broad}

\begin{figure}
\centering
\includegraphics[width=5.4 cm, angle=270]{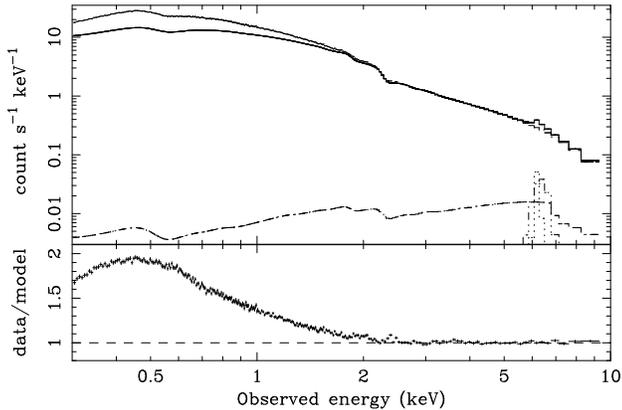}
\caption{
Broad-band EPIC pn spectrum (top panel) and data/model residuals
(bottom panel). The model spectrum (histogram) was fitted only above
$3.5$~keV and subsequently extrapolated to lower energies. The model
comprises a power-law plus reflection continuum and two Gaussian
emission lines (one narrow, one broad; see section~\ref{sect:iron}).
Galactic absorption was included in the model.
}
\label{fig:excess}
\end{figure}

The broad-band X-ray spectrum shown in Fig~\ref{fig:spectrum} is
that of a smooth, curving continuum, closely resembling that seen in
the X-ray spectra of some other Seyfert 1 galaxies including Ton S180
(Vaughan \et 2002), PKS 0558-504 (O'Brien \et 2001a) and Mrk 478
(Marshall \et 2003).   In order to gain a better view of the shape of
the broad-band spectrum, Fig~\ref{fig:excess} shows the  data/model
residuals when the simple model described in section~\ref{sect:iron}
was extrapolated to lower energies (compare with
Fig~\ref{fig:spectrum}).  Clearly there is a strong, smooth  soft
excess  which dominates at energies below $2$~keV. 
A similar pattern of residuals is seen in all three EPIC spectra.
This differs from
the soft excesses seen previously in e.g. Ton S180 and Mrk 478 in that
the excess appears to roll-over at energies below $\sim 0.5$~keV (at
least when compared to the underlying $\Gamma \approx 2$ hard X-ray
model).  This represents a `characteristic energy' of the soft
excess; the spectrum of the excess changes from steeper than $\Gamma
\approx 2$ above $\sim 0.5$~keV to a flatter spectrum at lower
energies (see also Fig.~\ref{fig:rgs_1}).  A similar roll-over in the
soft excess has previously been 
seen in a few Seyfert 1 galaxies (e.g. Mrk 359, O'Brien \et 2001b; NGC
4051, Collinge \et 2001).  This cannot be accounted for in terms of
excess neutral absorption, the curvature predicted by absorption 
does not match the
observed spectral form. The remainder of this section describes the
results of fitting  variety of continuum models to the data in an
attempt to account for the soft excess spectrum (see Vaughan \et
2002 for more discussion of these soft excess models).

For the purposes of this analysis the following assumptions were
made. The Galactic absorption was modelled using {\tt TBabs} with the
column density allowed to vary in the range $N_{\rm H} = 0.9-2.0
\times 10^{21}$~cm$^{-2}$ and there was assumed to be no absorption
intrinsic to \ark.   The models tested all accounted for the hard
X-ray continuum using a power-law (see section~\ref{sect:iron}).   In
addition a reflection spectrum was included (using the {\tt pexrav}
code unless stated otherwise) as was a two-component iron line
(modelled using the double Gaussian model discussed in
section~\ref{sect:iron}). The relative strength  of the reflection
($R$) as well as the normalizations of the two iron lines were
accepted as free parameters (but with the energies of the lines held
fixed).  This model acted as the `baseline' model to which various
continuum components were added in an attempt to model the soft
excess.  The differences in the spectral slopes between the pn and the
MOS  (see section~\ref{sect:iron}) can cause problems when attempting
to simultaneously fit broad-band continuum models to both
datasets. Therefore only the analysis performed on the pn data is
explicitly described, but at each stage a `sanity 
check' was performed by fitting to the MOS data independently.

The trial soft excess models were as follows:
($1$) an additional, soft power-law,
($2$) a break in the hard power-law,
($3$) two breaks in the power-law,
($4$) a blackbody,
($5$) two blackbodies, 
($6$) three blackbodies,
($7$) a bremsstrahlung thermal spectrum,
($8$) a thermal Comptonization spectrum (using the {\tt CompTT} code of Titarchuk 1994),
($9$) a disc blackbody spectrum (using the {\tt diskpn} code of Gierli\'{n}ski \et 1999),
($10$) thermal plasma (using the {\tt mekal} code of Kaastra 1992 and 
Liedahl, Osterheld \& Goldstein 1995);
($11$) ionized disc reflection and
($12$) ionized reflector plus bremsstrahlung
($13$) two separate ionized reflectors.
For models $11-13$ the {\tt pexrav} reflection model and 
broad Gaussian 
were replaced with the ionized reflection model of Ross \&
Fabian (1993) including convolution by a relativistic kernel (using
the {\tt Laor} code; Laor 1991) to model emission from 
an ionized accretion disc.
Table~\ref{tab:soft_fits} summarises the results of fitting these
different models to the broad-band EPIC pn spectrum.
In no case was the fit acceptable\footnote{For this reason
 the $\Delta \chi^2 = 2.7$ intervals 
 are not very meaningful as confidence intervals and
 so are not quoted.}
($P_{\rm rej}>99.9$ per cent). 

\begin{table}
\centering
\caption{
Fit statistics for various broad-band spectral models.
The models are listed as follows: 
pl = power-law; 
bknpo = broken power-law; 
2bknpo = doubly broken power-law; 
bb = blackbody; 
bremss = bremsstrahlung;
CompTT = thermal Comptonization;
diskpn = disc blackbody;
mekal = thermal plasma;
rf93 = ionized reflection (from Ross \& Fabian 1993).
The models are described in more detail in the text.
\label{tab:soft_fits}
}
\begin{tabular}{llr}
\hline
 & Model & $\chi^2/dof$ \\
\hline
1 & pl$+$pexrav$+$pl          & $3192.0 / 1522$ \\
2 & bknpo$+$pexrav            & $2921.5 / 1522$ \\
3 & 2bknpo$+$pexrav           & $1863.7 / 1520$ \\
4 & pl$+$pexrav$+$1$\times$bb & $2483.6 / 1522$ \\
5 & pl$+$pexrav$+$2$\times$bb & $1891.4 / 1520$ \\
6 & pl$+$pexrav$+$3$\times$bb & $1869.5 / 1518$ \\
7 & pl$+$pexrav$+$bremss      & $2227.7 / 1522$ \\
8 & pl$+$pexrav$+$CompTT      & $1944.2 / 1520$ \\
9 & pl$+$pexrav$+$diskpn      & $2199.0 / 1530$ \\
10 & pl$+$pexrav$+$mekal      & $4291.2 / 1530$ \\
11 & pl$+$rf93                & $2656.5 / 1522$ \\
12 & pl$+$2$\times$rf93       & $2130.4 / 1520$ \\
13 & pl$+$rf93$+$bremss       & $1898.2 / 1521$ \\
14 & pl$+$rf93$+$pl           & $2073.2 / 1521$ \\
\hline
\end{tabular}
\end{table}

\subsection{Power-law continua}

The two-slope power-law models $1$ (double power-law) and $2$ (singly broken
power-law) were unable to account for the
roll-over in the soft excess and produced very large data/model
residuals in the soft X-ray band. The power-law models were able to
fit down down to $\sim 0.6$~keV, the point at which the spectrum
starts to roll-over, but below this energy the source spectrum flattens
substantially. This is the case even after allowing for additional
absorption; including extra absorption in the model did not alter the
fit and could not account for the
roll-over in the soft X-ray spectrum. Thus the observed roll-over in
the soft excess must be intrinsic to the source and not an effect of
line-of-sight absorption.
However, the addition of a much flatter power-law slope at the lowest
energies greatly improved the fit (model $3$). In this
model the hard continuum has a slope of $\Gamma_{\rm hi} = 2.16$ down
to $E_{\rm hi} = 1.6$~keV where is steepened to $\Gamma_{\rm med} =
2.46$, accounting for the upturn in the spectrum.  Below $E_{\rm lo}
= 0.45$~keV the spectrum becomes very hard, with $\Gamma_{\rm lo} =
-0.46$, corresponding to the low energy roll-over in the soft excess.

\subsection{Thermal continua}

\begin{figure}
\centering
\includegraphics[width=5.2 cm, angle=270]{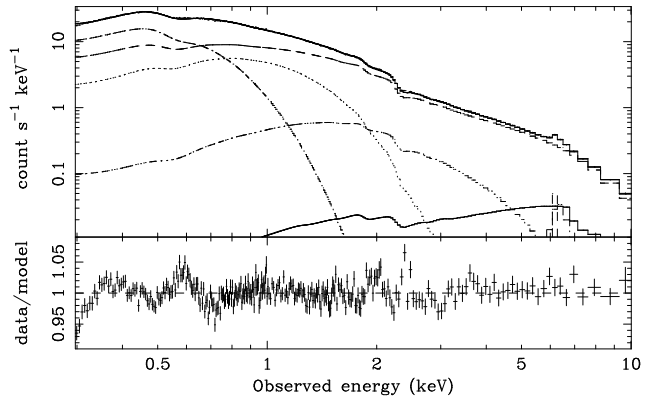}
\caption{
Broad-band EPIC pn spectrum (top panel) fitted with model (6) 
and data/model residuals (bottom panel). 
The model comprised a hard power-law plus 
reflection ({\tt pexrav}) and iron line emission.
The soft excess was accounted for with three blackbodies
spread over a range of temperatures.
The residual bump around $0.55$~keV coincides with the 
instrumental and Galactic oxygen edges (a similar pattern
of residuals was seen in the spectrum of the quasar 3c~273; Vaughan
\& Fabian 2004). 
}
\label{fig:3bb}
\end{figure}

The models based around thermal continuum emission (multiple
blackbodies, Comptonised blackbody, disc blackbody or bremsstrahlung)
provided comparably good fits and managed to account for  some degree
of the roll-over. Unsurprisingly, given the lack of emission lines in
the RGS spectrum, the  {\tt mekal} model gave a very poor fit using
solar  abundances. When allowed to be free the best-fitting abundance
was zero, which essentially reproduced a thermal bremsstrahlung
continuum.  Model $6$, shown in Fig~\ref{fig:3bb}, provided the best
fit of the thermal models, using  three blackbodies to model the soft
excess (with temperatures of $kT=102,~234,~673 $~eV).  The broad-band
($0.3-10$~keV) unabsorbed luminosity (predicted by this
three-blackbody model) was $L_{\rm X} \sim 2.8 \times 10^{44}$~erg
s$^{-1}$.

\subsection{Ionized reflection features}

The spectrum was also fitted with ionized reflection models
including relativistic smearing.
For all the reflection-based models the disc emission was very centrally
concentrated (e.g. $\rin < 1.4\rg$ for model $13$ and $\rin \approx
2.2\rg$ for model $14$). In fact, the best-fitting
emissivity index was often unreasonably steep ($q < -8$) and so the fits were
performed with $q=-6$ kept as a fixed parameter.
Furthermore, the steep emissivity meant the fitting was
insensitive to the $\rout$ parameter, this was therefore kept fixed at
$\rout = 400\rg$.  

It was not possible to model the entire soft excess  in terms of
emission from an ionized accretion disc (models $11$ and $12$). 
A single ionized reflector (model $11$) was unable to  fit the soft and hard 
bands simultaneously. The spectrum above $3$~keV was well fitted with
only weak ionization (section~\ref{sect:refln}), in which case the
reflector does not produce enough soft X-ray emission to account for
any of the soft excess. Allowing for an additional, independent
ionized reflector  (with much higher ionization; $\xi \sim 10^3$) still gave a poor
fit to the data  (model $12$). The primary reason for this failure
is that the highly ionized reflector, while producing a strong
soft excess, also produced very strong, broad spectral
features in the Fe K-band. Thus the bulk of the soft excess cannot be explained in
terms of ionized reflection (see also discussion in Vaughan \et 2002).

Ogle \et (2004) recently proposed that weak structures in the
soft excess of the bright Seyfert 1 galaxy NGC 4051 can be explained
by O~\textsc{viii} recombination emission (both the lines and
radiative recombination continuum, RRC) from a relativistic accretion
disc.  Models $13$ and $14$ work on a similar principle to the model
of Ogle \et (2004), in that the bulk of the soft excess is accounted
for using a continuum component (thermal bremsstrahlung or power-law)
and the relatively weak residual structures are fitted  using emission
from an ionized,  relativistic accretion disc).   These models gave
quite competitive fits to the data, superior to the ionized
reflection  models without an additional soft X-ray continuum
component.  In each case the disc was only mildly ionized ($\xi < 15$)
and so the fits were performed  with the ionization parameter kept
fixed at  $\xi = 10$.  At this low level of ionization 
the reflector contributed modest
equivalent width, broad emission features to the soft X-ray spectrum
(mainly from recombination to O$^{6+}$ and O$^{7+}$).  For example,
the equivalent widths of the emissions lines predicted by model $14$
were $\approx 36$~eV and $\approx 20$~eV for O~\textsc{vii} He$\alpha$
and O~\textsc{viii} Ly$\alpha$, respectively.

These models did provide reasonable
fits to the data (compared to the other models) and so it is at least
plausible  that some spectral structure in the soft excess of \ark\
could be accounted for by ionized disc emission.   Model $13$ accounts
for the majority of the soft excess using a $kT \approx 0.36$~keV
bremsstrahlung continuum whilst model $14$ used a $\Gamma \approx 3.6$
power-law.  Notably the bremsstrahlung continuum provided a better fit
to the soft X-ray spectrum than the power-law (see
Table~\ref{tab:soft_fits}).  However, a problem with both these models
is that the width of the observed Fe K$\alpha$ line in over-predicted
(see section~\ref{sect:iron}) by the strong gravitational redshift
on the reflection spectrum (implied by the small inner radius).

It should be noted that the Ross \& Fabian (1993) code was chosen to
model the soft X-ray  emission expected from an ionized disc, instead
of fitting the O~\textsc{viii} emission independently as was performed
by Ogle \et (2004). The model used by Ogle \et (2004) neglected a
number of effects that must be important in a realistic disc. Firstly
it did not account for emission lines from other species
(e.g. O~\textsc{vii}) nor from other ions (e.g. C, N, Fe) which should
also contribute to the soft X-ray spectrum (e.g. Ballantyne, Ross \&
Fabian 2002). Secondly it did not include the associated Compton
reflection continuum. Thirdly, the O~\textsc{viii} Lyman series line
ratios were calculated in the optically thin limit, whereas a
realistic accretion disc is expected to be optically thick to the
lines. This means that, before they can escape the disc, photons from
higher-order Lyman series lines (e.g. Ly$\beta$ etc.) will be degraded
into lower energy photons from other series and Ly$\alpha$
photons. Thus the higher-order Lyman lines should be strongly
suppressed, compared to the optically thin case, and the emission
dominated by the Ly$\alpha$ line. Therefore, a realistic accretion
disc spectrum would be expected to produce negligible emission in the
higher order Lyman lines. These effects are all included in the Ross
\& Fabian (1993) model.  (See also Dumont \et 2003 for more
details of line transfer effects in ionized discs.)

\subsection{Residual problems}

Even the best of the models discussed above gave a statistically
unacceptable fit to the data.  The most prominent fit residuals were
confined to three regions: $1.7-2.5$~keV, $0.5-0.7$~keV and $\ls
0.35$~keV.  The first set of residuals were almost certainly due to
instrumental Si and Au features (see e.g. Vaughan \& Fabian 2004).
The $0.5-0.7$~keV residuals could plausibly due due to low level,
broad emission structures (see above), but they coincide with the
instrumental and Galactic oxygen edges (at $0.543$~keV). The RGS data
showed that there were no strong, narrow emission/absorption lines at
$\sim 0.55-0.60$~keV ($20.7-22.5$~\AA). The  MOS spectral residuals
showed noticeably less structure than the pn residuals in this region.
Thus, given the present calibration uncertainties, it is not yet
possible to unambiguously  assess the shape and significance of these
residuals.  Apart from these two specifics, the
remaining residuals were smaller than $\ls 5$ per cent and so the best
of the above models could be considered reasonable fits to the data
(even though the over-all $\chi^2$ values were poor).  The slight
down-turn in the residuals below $0.35$~keV (e.g. Fig.~\ref{fig:3bb})
may  indicate that even for the best-fitting models the spectral
curvature was not enough to explain the spectral roll-over.

% --------------------------------------------------------------------------
% --------------------------------------------------------------------------
% --------------------------------------------------------------------------

\section{\xte\ data}
\label{sect:xte}

\begin{figure}
\centering
\includegraphics[width=6.40 cm, angle=270]{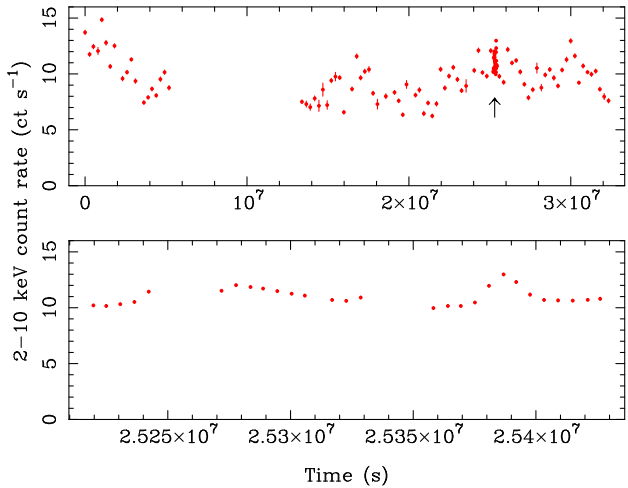}
\caption{
Top panel: 
\xte\ light curve of \ark\ covering the period 1998 Feb 24 to 1999 Mar
14. The arrow indicates the period of intensive
monitoring (1998 Dec 15--18).
Bottom panel: light curve obtained from the intensive monitoring.
}
\label{fig:xte_lc}
\end{figure}

\ark\ has been monitored by \xte\ for several years.
Fig~\ref{fig:xte_lc} shows a section of the \xte\
light curve spanning approximately one year which clearly
reveals the source is variable but over timescales longer than sampled
by \xmm. Marshall \et (2004) discuss the \xte\ monitoring in more
detail.

Over the period 1998 Dec 15--18 \xte\ observed  \ark\ rather more
intensively, resulting in $67.6$~ks of useful data from 3 PCUs. These
data were reduced using standard methods  (see e.g. Markowitz, Edelson
\& Vaughan 2003) and a time averaged $3-20$~keV spectrum was extracted.
The spectrum clearly reveals the iron line and also suggests
the presence of a reflected continuum. 

\begin{figure}
\centering
\includegraphics[width=6.40 cm, angle=270]{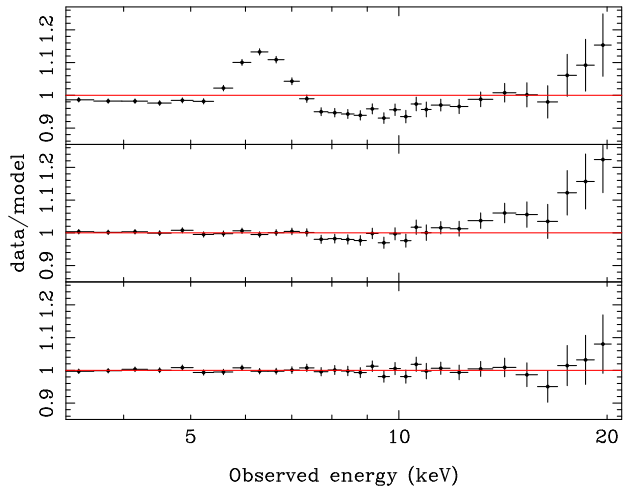}
\caption{
\xte\ spectrum of \ark\ taken from the period of intensive
monitoring (1998 Dec 15--18).
The three panels show the data/model residuals using the
following three models: a simple power-law (top), 
and power-law plus Gaussian (middle) and a power-law plus Gaussian
and reflection (bottom).
All three models included Galactic absorption.
}
\label{fig:xte_spec}
\end{figure}

Three simple models were compared to the data as detailed in
Table~\ref{tab:xte}. These were ($1$) a power-law, ($2$) a
power-law plus Gaussian and ($3$) a power-law plus Gaussian
and reflection continuum. The reflection continuum was
modelled using the {\tt pexrav} code.
Galactic absorption was included in all three models.
The iron line was modelled using a single broad Gaussian, which was
adequate given the low spectral resolution of the \xte\ PCA.
Fig~\ref{fig:xte_spec} shows the residuals from each model.
The simple power-law provided an unacceptable fit.
The fit became acceptable once a Gaussian was added, but 
was significantly improved by the addition of a reflection continuum. 
Although these \xte\ data were taken almost five years
prior to the \xmm\ observation, 
the \xte\ spectrum lends weight to the \xmm\ results by
confirming the presence of an iron line and supporting the
existence of a weak ($R \sim 0.5$) reflection continuum.

\begin{table}
\centering
\caption{Best-fitting parameters for models of the \xte\ spectrum.
\label{tab:xte}}
\begin{tabular}{llr}
\hline
Model & Parameter & $\chi^{2}/dof$ \\
\hline
power-law     &  $\Gamma=1.91$       & $631.4/44$   \\
 \\
power-law +   & $\Gamma=1.93\pm0.01$ & $51.9/41$    \\
Gaussian      & $E=6.47\pm0.04$~keV  &              \\
              & $\sigma=260\pm90$~eV &              \\
              & $EW=217\pm20$~eV     &              \\
 \\
power-law +   & $\Gamma=2.01\pm0.03$ & $24.0/40$    \\
Gaussian +    & $E=6.50\pm0.05$~keV  &              \\
              & $\sigma=328\pm90$~eV &              \\
              & $EW=224\pm22$~eV     &              \\
reflection    & $R=0.50\pm0.19$      &              \\
\hline
\end{tabular}
\end{table}

Very similar results were obtained from an analysis of the
time average spectrum of the long timescale monitoring (excluding the
1998 Dec 15-18 intensive monitoring data). 

% --------------------------------------------------------------------------
% --------------------------------------------------------------------------
% --------------------------------------------------------------------------

\section{Discussion}
\label{sect:disco}

\subsection{The X-ray properties of \ark}

This paper describes the results of a  long \xmm\ observation of the
luminous Seyfert 1 galaxy Arakelian 120. The X-ray emission from the
source was only weakly variable on the short timescales probed by \xmm\
(section~\ref{sect:timing}) but archival \xte\ data revealed `normal'
Seyfert 1 variability traits on longer timescales
(section~\ref{sect:xte}; cf. Edelson \& Nandra 1999; Uttley \et 2002;
Markowitz \et 2003b).  The X-ray spectrum showed a notable absence of
warm absorption features (section~\ref{sect:rgs}).
The lack of absorption means that these data represent the `bare'
X-ray emission spectrum of a fairly typical Seyfert 1. 

The spectrum above $\sim 3$~keV can be explained using a fairly
conventional spectral model (sections~\ref{sect:iron} and
\ref{sect:refln}) comprising a power-law continuum ($\Gamma \approx
2$) plus Doppler broadened emission from the surface of a
weakly ionized reflector (with relative reflection strength $R
\approx 0.5$). The \xte\ data support this model
(section~\ref{sect:xte}).  The spectrum below $\sim 3$~keV becomes
dominated by a steep, smooth, broad soft excess component which
appears to peak at $\sim 0.5$~keV.

\subsection{The iron line of \ark}

The results of the spectral fitting indicated an iron emission line that
is the composite of a weak, narrow line originating in distant
material and a stronger line showing significant Doppler broadening.

A narrow, neutral iron line appears to be ubiquitous in Seyfert 1
galaxies (e.g. Yaqoob, George \& Turner 2002; Page \et 2004b; Yaqoob
\& Padmanabhan 2004) and may have an original in the optical
broad-line region or the putative molecular torus (see discussion in
Yaqoob \et 2001; Reeves \et 2004). The lack of neutral absorption
intrinsic to \ark\ (section~\ref{sect:rgs}) confined this
material to lie out of the line of sight, which in turn implies the
covering fraction of the line emitting material must be below unity.
The relative strength of the narrow line ($EW \sim 40$~eV) further suggests
the sky covering fraction of the material is small ($f_{\rm C} \ls
0.5$; Nandra \& George 1994) or the optical depth is small ($\tau \ls
0.1$; Leahy \& Creighton 1993) or both.

The broad component to the line has a velocity width  $FWHM \sim
3\times 10^4$ km s$^{-1}$, far broader than the broad optical lines
(e.g. $FWHM({\rm H}\beta) \approx 5800$~km s$^{-1}$; Wandel \et 1999).
However, there is no requirement for the line emitting region to
extend into the region of strong gravity about the black hole ($\ls
20$~\rg) which would produce an asymmetric redward tail  on the line
profile (section~\ref{sect:iron}). The best-fitting parameters of the
disc line model are slightly unusual; the inclination angle is
high ($i > 73 \deg$), the inner radius is rather large
($r_{\rm in} \sim 140 r_{\rm g}$) and the rest-frame energy is
unusual ($E \approx 6.56$~keV). 
One plausible origin for this line is a weakly ionized accretion
disc. The fact that the inner radius is greater than $6r_{\rm g}$
could mean the disc is truncated. However, the unusual energy of  the
line (corresponding to mildly ionized iron) might allow for a simpler
alternative. If the disc survives down to the ISCO but rapidly becomes
ionized with decreasing radius then, depending on the detailed
ionization structure, the innermost regions may produce
little observable line emission (see e.g. Ross, Fabian \& Young 1999).

\begin{figure}
\centering
\includegraphics[width=6.0 cm, angle=270]{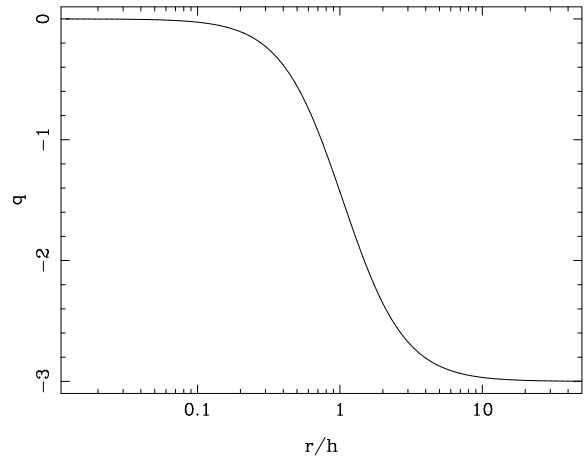}
\caption{
The local power-law slope of the 
emissivity function  ($J(r) \sim r^q$) for a lamppost geometry.
}
\label{fig:emissivity}
\end{figure}

However, as demonstrated by Fig.~\ref{fig:contour} there is sufficient
leverage in the fit to allow for a smaller inner radius ($r_{\rm in}
\ls 100 r_{\rm g}$), and also lower inclination angle, provided the
emissivity law is quite flat ($q \approx -2$). 
As mentioned in section~\ref{sect:iron} a good fit can be obtained
with $r_{\rm in}=6\rg$ and $i \approx 30 \deg $ provided that 
there are two co-existing Fe K$\alpha$ emission lines arising in the
disc (at energies of
$6.4$ and $6.8$~keV) and no intrinsically narrow component. This may
be feasible if the disc is clumpy 
or inhomogeneous (Ballantyne, Turner \& Blaes 2004).
The required 
flat emissivity can be produced if the disc is illuminated from a great
height. Figure~\ref{fig:emissivity} shows the effect of altering $r/h$
in a `lamppost' geometry [i.e. the disc was assumed to be flat and
illuminated by a central point source raised by a height $h$, in
which case the emissivity law is given by $J(r) \propto
h/(r^2+h^2)^{3/2}$]. This demonstrates that if the X-illumination
is provided from a height $h \sim 100 r_{\rm g}$ the emissivity will
be quite flat out to similarly large radii. 
Illumination from a large height would require a non-standard geometry
for the X-ray emitting region, one possibility is the `aborted jet' model
of Ghisellini, Harrdt \& Matt (2004).

\subsection{\ark\ as a `bare' Seyfert 1}

These new X-ray observations showed \ark\ to posses
no evidence of an X-ray warm absorber and placed upper limits on the ionic column
densities that are substantially  lower than those of more typical,
absorbed Seyfert 1s. For
example, NGC 3783 shows O~\textsc{vii} and O~\textsc{viii} absorption
with column densities two orders of magnitude higher (Kaspi \et
2002). Kaastra \et (2000) observed NGC 5548 and measured
C~\textsc{vi}, N~\textsc{vi}, O~\textsc{vii} and O~\textsc{viii}
absorption lines with corresponding ionic column densities an
order of magnitude higher than the limits obtained for \ark.  A last
counter example is IRAS 13349+2438 (Sako \et 2001) which shows
absorption lines  of C~\textsc{vi}, N~\textsc{vi}, O~\textsc{vii},
O~\textsc{viii}, Ne~\textsc{ix} and Ne~\textsc{x}, with column
densities an order of magnitude higher than \ark.  

These X-ray observations therefore confirm that \ark\ is indeed a
`bare' Seyfert 1 nucleus, as suspected based on its broad band
spectral energy distribution (SED; Edelson \& Malkan 1986; Ward \et
1987) and the lack of ultraviolet absorption seen in \hst/FOS spectrum
(Crenshaw \et 1999). 
Other ultraviolet observations have been made with the  Goddard
High-Resolution Spectrograph (Penton, Stocke \& Shull 2000) and with
\fuse\ (Wakker \et 2003). Using these data Wakker \et (2003) showed
that the line-of-sight through the Galaxy towards \ark\ possesses a
very low column of O~\textsc{vi}.  A preliminary examination of these
data suggest there may be weak traces of ionized absorption intrinsic
to \ark.

\subsection{Other bare Seyferts}

Shortly after the original discovery of the X-ray warm absorber (Halpern
1984), this phenomenon was found to be common in Seyfert 1
galaxies. Low resolution spectra from \exosat\ and \ginga\
suggested $\sim 50$ per cent of Seyfert 1s showed evidence for warm
absorption (Turner \& Pounds 1989; Nandra \& Pounds 1994).  Subsequent
observations with the better spectral resolution afforded by \asca\
suggested the incidence of warm absorbers was possibly higher ($\sim
50-70$ per cent; Reynolds 1997; George \et 1998).  Similar results
were found for the ultraviolet absorber based on  \hst\ FOS spectra
(Crenshaw \et 1999).

The increased sensitivity and resolution offered by \xmm\ and
\chandra\ have allowed for even more sensitive searches for
absorption. These have confirmed that there remains a significant
population of Seyfert 1s that lack a strong X-ray warm absorber.  In
addition to \ark,  other Seyfert 1s that show a distinct lack of warm
absorption  in both their \hst\ FOS ultraviolet and \xmm/\chandra\
X-ray spectra include Mrk 478 (Marshall \et  2003a),  Mrk 335 (Gondoin
\et 2002), Fairall 9 (Gondoin \et 2001) and Mrk 205 (Reeves \et 2001).
All four of these were in the sample of Crenshaw \et (1999)  and
classified as having no intrinsic ultraviolet warm absorber.   Other
objects known to lack X-ray absorbers include  PKS 0558--504 (O'Brien
\et 2001a), MCG--2-58-22 (Weaver \et 1995) and Ton S180\footnote{
There is some debate about Ton S180.  R\'{o}\.{z}a\'{n}ska \et (2004)
re-examined the \chandra\ LETGS observation and claim to have found
several resonance absorption lines.  The lines are very weak and in
most cases of borderline significance (many have equivalent widths
consistent with zero).  Furthermore, the spectral analysis was
complicated by the presence of contaminant on the \chandra\ ACIS which
affects the LETGS/ACIS calibration (see Marshall \et 2003b).  Thus the
existence of an X-ray warm absorber in Ton S180 is still not well
established and in any case must be extremely weak.  } (Turner \et
2001b; Vaughan \et 2002).

These objects are probably not completely without ionized
absorption systems.
Kriss (2002) reported weak absorption by the O~\textsc{vi} $\lambda
\lambda 1032, 1038$~\A\ resonance doublet in  both Ton S180 and Mrk
478 based on high resolution \fuse\ data (see also Turner \et 2002 for
the \fuse\ observation of Ton S180). Such absorption is too weak to
have been detected in the Crenshaw \et (1999) ultraviolet survey. The
corresponding X-ray warm absorbers in these objects may be present but
so weak as to have a negligible effect on the available data.

The reason for the lack of an X-ray warm absorber is unclear.  The
most obvious explanation is that overall column density  is lower in
these objects (by at least an order of magnitude) compared to  more
typical, absorbed Seyferts. However, it is also plausible that a
similar column of ionized gas exists but is either too highly ionized
to show significant spectral features or lies out of the line of sight
(requiring a covering fraction less than unity).  A detailed survey
comparing the X-ray and ultraviolet emission/absorption line spectra
of a large sample of bright Seyferts may be able to answer this
question (see discussions in Crenshaw \et 2003).
A more speculative solution, recently proposed by Gierli\'{n}ski \&
Done (2004), is that objects that lack narrow absorption lines may be
dominated by a deep, broad absorption trough produced by an absorption
system with such high dispersion velocity that no individual
lines can be resolved. This model would require the underlying power-law
continuum to be rather steep, which then makes explaining the
upturn above $\sim 10$~keV more difficult to explain.

\subsection{How common are soft X-ray excesses?}

These observations have clearly revealed a luminous soft excess in
\ark, in the sense that an extrapolation of the hard ($\gs 3$~keV)
continuum into the soft X-ray band revealed a very significant upturn.
Steep soft X-ray spectra were first seen in Seyferts using \heao\ data
(Pravdo \et 1981)   and subsequently seen as excesses over the hard
power-law  by \exosat\ (Arnaud \et 1985) and \einstein\ (Bechtold \et
1987).  Early soft X-ray surveys (e.g. the \exosat\ survey of Turner
\& Pounds 1989) suggested that $\gs 50$ per cent of all Seyferts
possessed a soft excess component.  The \rosat\ survey of Walter \&
Fink (1993) also suggested a high incidence of soft excesses.  These
numbers were highly uncertain, however,  because many of the sample
members were absorbed.  Complex soft X-ray absorption can mask, or in
some cases even mimic, the steeper soft X-ray spectrum indicative of a
soft excess.

The recent \xmm\ and \chandra\ observations would seem to
indicate that all Seyfert 1s without strong X-ray warm absorption
do show a strong soft excess.  However, many of the well-known examples
(e.g. Ton S180, PKS 0558-504, Mrk 478, NGC 4051 and Mrk 359) are
narrow-line Seyfert 1s (NLS1s; Osterbrock \& Pogge 1985). NLS1s are a
subclass of Seyfert 1s defined by their narrow permitted optical lines
($FWHM({\rm H}\beta) \ls 2000$~km s$^{-1}$) but noted for their often
exceptionally steep soft X-ray spectra (Boller \et 1996; Laor \et
1997; Vaughan \et 1999; Leighly 1999). Until recently it was possible
that the soft excess appeared ubiquitous in unabsorbed Seyferts only
because many of the well-studied examples were NLS1s (i.e. the sample
of unabsorbed Seyferts was biased towards the soft NLS1s).

\ark\ is an interesting counter example, being both unabsorbed and a
`normal,' broad-line Seyfert 1 (BLS1) with  $FWHM({\rm H}\beta)
\approx 5800$~km s$^{-1}$ (Wandel \et 1999). Yet it too shows a strong
soft excess. Other notable BLS1s that also lack absorption include
Mrk 205 (Reeves \et 2001), Mrk 335  (Gondoin \et 2002) and Fairall 9
(Gondoin \et 2001),  all of which showed soft excesses in their \xmm\
observations.  Although a complete and unbiased survey of the soft
X-ray spectra of Seyfert 1s has yet to be conducted, it does seem
highly plausible that all unabsorbed Seyfert 1s (whether NLS1 or BLS1)
possess a soft excess. Assuming that there is no fundamental
difference in the underlying X-ray continuum spectra of absorbed and
unabsorbed Seyfert 1s then implies that soft X-ray excesses are
ubiquitous to Seyfert 1s. The mini-survey of \xmm\ spectra by Pounds
\& Reeves (2002) showed that in all six Seyfert 1s they studied, the
soft X-ray spectrum ($\ls 0.5$~keV) was always  $\gs 50$ per cent
higher than an extrapolation of the hard X-ray power-law would
predict.  Furthermore, detailed studies of Seyfert 1s with complex,
strong X-ray warm absorbers often conclude that an additional soft X-ray
emission component is required behind the warm absorber (e.g. Collinge
\et 2001; Netzer \et 2003;  Blustin \et 2003).

This therefore underlines the need to understand the soft excess as a
common (perhaps ubiquitous) and significant contributor to the
luminosity of Seyferts.  In the case of \ark\ the best-fitting models
are a doubly-broken power-law with an anomalously flat slope below
$0.5$~keV, or multiple, soft blackbodies. The blackbody origin is
difficult to explain as the temperatures are far too
high to correspond to any standard accretion disc. The expected
temperature for the inner region ($\approx 6r_{\rm g}$) of a
geometrically thin, optically thick disc about a $2
\times10^{8}$~\Msun\ black hole is $kT \sim 11$~eV if radiating at
$L/L_{\rm Edd}=0.1$ and $kT \sim 20$~eV if radiating at $L/L_{\rm
Edd}=1$. This emission should therefore not contribute to the observed
X-ray spectrum. The temperatures of the best-fitting blackbody
components were at at least an order of magnitude higher than this. In
addition, the size of the emission region implied by these high
temperatures is far too small ($ r_{\rm BB} \ll r_{\rm g}$).  The
alternative models tested (reflection, disc blackbodies and
bremsstrahlung) all produced the wrong spectral shape.  At present
there is no single model that can account for the known properties of
the soft excess (see also  the discussion in Vaughan \et 2002).

% --------------------------------------------------------------------------
% --------------------------------------------------------------------------
% --------------------------------------------------------------------------

\section*{Acknowledgements}

Based on observations obtained with \xmm, an ESA science mission with
instruments and contributions directly funded by ESA Member States and
the USA (NASA).  We would like to thank the SOC and SSC teams for
making possible the observations and data analysis, and an anonymous
referee for a helpful report.  SV acknowledges
financial support from PPARC.  This research has made use of the
NASA/IPAC Extragalactic Database (NED) which is operated by the Jet
Propulsion Laboratory, California Institute of Technology, under
contract with the National Aeronautics and Space Administration.
This research also made use of observations made with the NASA/ESA
Hubble Space Telescope, obtained through the data archive at the Space
Telescope Science Institute. STScI is operated by the Association of
Universities for Research in Astronomy, Inc. under NASA contract NAS
5-26555.

\bsp
\label{lastpage}
\end{document}